# The bounds of the set of equivalent resistances of n equal resistors combined in series and in parallel

Sameen Ahmed KHAN

(rohelakhan@yahoo.com, http://SameenAhmedKhan.webs.com/)

Engineering Department,
Salalah College of Technology (SCOT)
Salalah, **Sultanate of Oman**

**Abstract:**
The order of the set of equivalent resistances, $A(n)$ of n equal resistors combined in series and in parallel has been traditionally addressed computationally, for n up to 22. For larger n there have been constraints of computer memory. Here, we present an analytical approach using the Farey sequence with Fibonacci numbers as its argument. The approximate formula, $A(n) \sim 2.55^n$, obtained from the computational data up to n = 22 is consistent with the strict upper bound, $A(n) \sim 2.618^n$, presented here. It is further shown that the Farey sequence approach, developed for the $A(n)$ is applicable to configurations other than the series and/or parallel, namely the bridge circuits and non-planar circuits. Expressions describing set theoretic relations among the sets $A(n)$ are presented in detail. For completeness, programs to generate the various integer sequences occurring in this study, using the symbolic computer language MATHEMATCA, are also presented.

## Introduction

In an introductory physics course one finds exercises such as: *Find all the resistances that can be realized using three equal resistors in various combinations* [1]. The 4 possible solutions (there are six solutions, if the exercise is to use *3 or fewer* equal resistors) are:

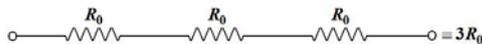

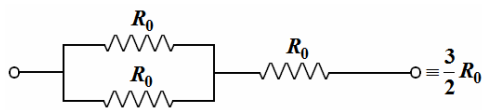

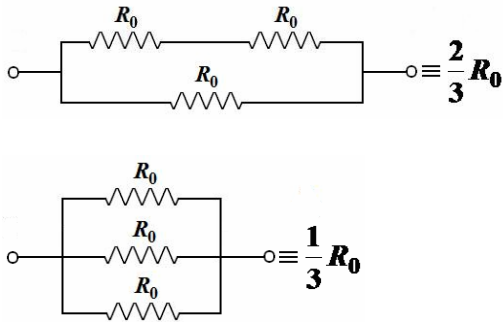

**Fig.-1. Circuit configurations for 3 resistors**

The solutions are proportional to the unit resistance $R_0$. This unit resistance can be set to unity without any loss of generality. We continue the exercise with 4 resistors and find 9 equivalent resistances, but ten different configurations. The following two configurations have the same equivalent resistance.

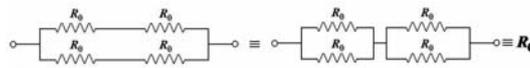

**Fig.-2. Equivalent Circuits in the case of four resistors**

At this point, we note that different configurations can give rise to the same equivalent resistance. This scenario of different configurations giving the same equivalent resistance persists for 4 and more resistors. Next, we analyze the case of five resistors. One possible configuration is the bridge network [2], whose equivalent resistance for equal resistors is 1.

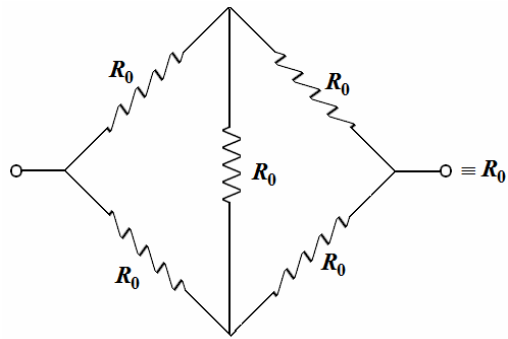

**Fig.-3. Bridge Circuit**

Rest of the configurations, use exclusively series and/or parallel combination and result in 22 equivalent resistances. The number of configurations is much larger than the number of equivalent resistances. For n = 22, the number of configurations is 35 times more than the number of equivalent resistances. Far arbitrary larger number of resistors the ratio between the number of configurations and the number of equivalent resistances diverges.

In this article we are primarily interested in counting the set of equivalent resistances using equal resistors in series and/or parallel combinations. The task is to compute A(n), the order of the set of equivalent resistance using all the n resistors of equal value. Even with these assumptions the order of the set of equivalent resistances grows rapidly and we have for n = 1, 2, 3, …, A(n) = 1, 2, 4, 9, 22, 53, 131, 337, …, respectively. The problem for n up to 16 has been addressed computationally [3]. In this article we shall address the question analytically and provide an upper bound for A(n). The approximate formula, $A(n) \sim 2.55^n$, obtained from the numerical data up to n = 16 in [3] is consistent with the analytical result, $A(n) \sim 2.618^n$ presented here. It is to be noted that in [3], $2.55^n$ was obtained by fitting a straight line to ln(A(n)) against n for the points n = 6 to n = 16. Direct ratios of A(n + 1)/A(n) lead to 2.53 for n = 15; the relevant details are in Table-1. We shall initially consider the case of using all the n resistors; then extend it to the case of n *or fewer* resistors (at most n resistors). It shall also be shown that the framework developed for the A(n) is very much applicable to configurations involving bridge circuits and non-planar circuits. For completeness, certain technical details are presented in appendices.

Details pertaining to circuit analysis are presented in Appendix-A. The same appendix also covers the extended sets obtained by incorporating the bridge configurations into the sets containing the series and parallel configurations. Proofs of two theorems used in the derivation of the strict upper bound are presented in Appendix-B and Appendix-C respectively. Mathematical notes on the Fibonacci numbers and the Haros-Farey sequence are presented in Appendix-D. Appendix-E is dedicated to the set theoretic relations among A(i). The relevant computational programs to generate the various integer sequences occurring in this study, using the symbolic computer language MATHEMATCA, are presented in Appendix-F.

The sequence A(n) is the first sequence occurring in this article to be followed by several more. We shall cite the various integer sequences occurring in this study, by the unique identity assigned to each of them in *The On-Line Encyclopedia of Integer Sequences* (OEIS), created and maintained by Neil Sloane [4]; and list them towards the end of the bibliography. For instance the sequence A(n) is identified by A048211 and OEIS has six additional terms up to n = 22, where as [3] contains the first 16 terms. A short note on the integer sequences and the OEIS is presented in Appendix-G.

**Results and Analysis**
Let $R_0$ be the value of the *n* equal resistors being used. The net resistance of all the configurations is proportional to $R_0$. So, we can set its value to be unity. The proportionality constant is a rational number (say a/b; with a and b being natural numbers, a/b is in its reduced form) depending on the configuration. The value of a/b ranges from *1/n* (for all the n resistors in parallel configuration) to *n* (for all the n resistors in series configuration). The task is to calculate the set of values of a/b obtained for all conceivable configurations. The set of values of a/b, for n up to six are:

$$A(1) = 1 : \left\{ n = 1 : 1 \right\},$$

$$A(2) = 2 : \left\{ n = 2 : \frac{1}{2}, 2 \right\},$$

$$A(3) = 4 : \left\{ n = 3 : \frac{1}{3}, \frac{2}{3}, \frac{3}{2}, 3 \right\},$$

$$A(4) = 9 : \left\{ n = 4 : \frac{1}{4}, \frac{2}{5}, \frac{3}{5}, \frac{3}{4}, 1, \frac{4}{3}, \frac{5}{3}, \frac{5}{2}, 4 \right\},$$

$$A(5) = 22 : \left\{ n = 5 : \frac{1}{5}, \frac{2}{7}, \frac{3}{8}, \frac{3}{7}, \frac{1}{2}, \frac{4}{7}, \frac{5}{8}, \frac{5}{7}, \frac{4}{5}, \frac{5}{6}, \frac{6}{7}, \frac{7}{6}, \frac{6}{5}, \frac{5}{4}, \frac{7}{5}, \frac{8}{5}, \frac{7}{4}, 2, \frac{7}{3}, \frac{8}{3}, \frac{7}{2}, 5 \right\},$$

$$A(6) = 53 : \left\{ n = 6 : \begin{array}{l} \frac{1}{6}, \frac{2}{9}, \frac{3}{11}, \frac{3}{10}, \frac{1}{3}, \frac{4}{11}, \frac{5}{13}, \frac{5}{12}, \frac{4}{9}, \frac{5}{11}, \frac{6}{13}, \frac{7}{13}, \frac{6}{11}, \\ \frac{5}{9}, \frac{7}{12}, \frac{8}{13}, \frac{7}{11}, \frac{2}{3}, \frac{7}{10}, \frac{8}{11}, \frac{3}{4}, \frac{7}{9}, \frac{4}{5}, \frac{5}{6}, \frac{9}{10}, \frac{10}{11}, 1, \\ \frac{11}{10}, \frac{10}{9}, \frac{6}{5}, \frac{5}{4}, \frac{9}{7}, \frac{4}{3}, \frac{11}{8}, \frac{10}{7}, \frac{3}{2}, \frac{11}{7}, \frac{13}{8}, \frac{12}{7}, \frac{9}{5}, \\ \frac{11}{6}, \frac{13}{7}, \frac{13}{6}, \frac{11}{5}, \frac{9}{4}, \frac{12}{5}, \frac{13}{5}, \frac{11}{4}, 3, \frac{10}{3}, \frac{11}{3}, \frac{9}{2}, 6. \end{array} \right\}.$$

Throughout this article, we shall use the same symbols, A(n), B(n), etc., to denote the set and its order respectively. From the above sets, we make the following observations. Each value of a/b in a given set occurs in a reciprocal-pair of a/b and b/a respectively (1 being its own partner). The largest value of a and b in a given set is equal to Fibonacci(n + 1), the (n + 1)[th] term in the Fibonacci sequence[5-6, A000045] (for details of the Fibonacci sequence, see the Appendix-D). We shall shortly prove that these observations are true for all n. Two theorems summarizing the occurrence of equivalent resistances in reciprocal pairs and the largest values of the numerator/denominator bounded by the Fibonacci respectively, shall be used to derive the upper bound of A(n). It is also to be noted that the sets A(n) of higher order do not necessarily contain the complete sets of lower orders. For example 2/3 is present in the set A(3), but it is not present in the sets A(4) and A(5). The element 1 belongs to all sets A(n), except the three sets, A(2), A(3) and A(5). Consequently, all A(n) except these three are odd. Such set theoretic statements and interrelations among A(i) are presented along with their proofs in Appendix-E.

A circuit constructed from n resistors, by adding one resistor in series or in parallel at a time leads to (n − 1) connections, which can be either in series or in parallel (two possibilities per connection). One can have additional configurations when one adds more than one resistor at a time. This gives us the strict inequality

$$\frac{1}{2} \times 2^n < A(n) .$$

We choose to write half as a multiplicative factor in order to point to the trend, $2^n$ for n resistors. The lower bound thus obtained is off the mark. The order of the set $A(n + 1)$ can be estimated from the set $A(n)$. Treating the elements of $A(n)$ as single blocks the $(n + 1)^{th}$ resistor can be added either in series or in parallel. We call these two sets as series set and parallel set and denoted them by $1SA(n)$ and $1PA(n)$ respectively. One can also add the $(n + 1)^{th}$ resistor somewhere within the $A(n)$ blocks, and we call this set as the cross set and denote it by $1 \otimes A(n)$. The series and the parallel sets each have exactly $A(n)$ number of configurations and the same number of equivalent resistances. All the elements of the parallel set are strictly less than 1 and that of the series set are strictly greater than 1. These two disjoint sets contribute $2A(n)$ number of elements to $A(n + 1)$ and are the source of $2^n$. The cross set is not straightforward, as it is generated by placing the $(n + 1)^{th}$ resistor anywhere within the blocks of $A(n)$. It is the source of all extra configurations, which do not necessarily result in new equivalent resistances. The cross set has at least $A(n − 1)$ elements, since $A(n)$ has $A(n − 1)$ connections corresponding to $1 \otimes A(n − 1)$. This argument works for $n \geq 6$, leading to the inequality

$$A(n + 1) > 2A(n) + A(n − 1) .$$

Solving, the above recurrence equation (leading to the quadratic equation $\lambda^2 = 2\lambda + 1$), we obtain $\lambda = (1 + \sqrt{2}) = 2.4142$, leading to the lower bound

$$\frac{1}{4} \times (1 + \sqrt{2})^n < A(n) .$$

This lower bound, $A(n) < 2.41^n$ based on an approximate recurrence relation is consistent with the numerical result $A(n) \sim 2.55^n$, obtained from the numerical data up to n = 16 in [3] and n = 22 in [A048211]. The series and parallel sets are straightforward: they have the same number of configurations as the number of equivalent resistances; are disjoint and distributed on either side of the element 1; lead to the relation $2^n$; etc.,. The cross set is not so well understood and hence, it does not appear feasible to get a good count of the set of equivalent resistances, using combinatorial arguments.

The strict upper bound of $A(n)$ is based on the following two theorems.

*Theorem-1* (*Reciprocal Theorem*):
In any circuit constructed from n equal resistors (of value $R_0$) in series and/or parallel combination has an equivalent resistance $(a/b)R_0$, then the configuration obtained by changing all series connection to parallel and parallel connections to series respectively, results in a configuration, whose equivalent resistance is $(b/a)R_0$.

*Theorem-2* (*Bound of a and b Theorem*):

Any circuit constructed from n equal resistors (of value $R_0$) in series and/or parallel combination has an equivalent resistance $(a/b)R_0$, such that the largest possible values of a and b are bounded by the $Fib(n + 1)$, the $(n + 1)^{th}$ term in the Fibonacci sequence. This largest value $Fib(n + 1)$ is obtained by taking the combinations in series and parallel incrementally.

The proofs of the above theorems are presented in Appendix-B and Appendix-C respectively. The reciprocal theorem states that resistances in the set A(n) always occur in the pairs a/b and b/a, one less than 1 and the other greater than 1. So, it suffices to count the number of configurations, whose equivalent resistance is less than 1. Theorem-2 fixes the bound on the values of a and b. So, the problem of deriving A(n) translates to counting the number of 'relevant' proper fractions whose denominators are bounded by m = $Fib(n + 1)$. Farey sequence provides the most exhaustive set of fractions in the interval [0, 1], whose denominator are less than or equal to a given natural number m [7, 8, A005728]. For details of the Farey sequence, see Appendix-D. The set of proper fractions in the set for A(n) is bounded in the subinterval I = [1/n, 1], which is a subinterval of [0, 1]. Note, that the resistance, 1/n is obtained by taking all the n resistors in parallel. The length of the interval, I is |I| = 1 − 1/n. For large n this is almost unity. Taking into account that all elements of A(n) have a reciprocal pair except 1, and the fact that 1 is included in the Farey sequence (1 gets counted twice); we have the expression

$$G(n) = 2Farey(m; I) - 1 = 2Farey\big(Fib(n+1); I\big) - 1.$$

G(n), by construction is the grand set (superset) containing the fractions from Farey(Fib(n + 1)) in the interval [1/n, 1] along with their reciprocals [A176502]. Appendix-D contains additional details regarding the construction of the grand set G(n). 1 is also an element of G(n). In other words G(n) contains all rational numbers of the form a/b such that both a and b are bounded by $Fib(n + 1)$. Since the Farey sequence is exhaustive, the set G(n) is also exhaustive. This leads to the strict upper bound

$$A(n) < G(n) = 2Farey\big(Fib(n+1); I\big) - 1.$$

Ignoring the −1 in the above expression, and using the asymptotic relation for Farey(m) and the closed form expression for Fib(n + 1) we have

$$G(n) \sim 2 \times |I| \frac{3}{\pi^2} m^2 \sim (1 - \frac{1}{n}) \frac{6}{\pi^2} \left(\frac{\phi^{n+1}}{\sqrt{5}}\right)^2 = (1 - \frac{1}{n}) \frac{6\phi^2}{5\pi^2} \phi^{2n} = (1 - \frac{1}{n}) 0.318 \times (2.618)^n,$$

where $\phi = (1 + \sqrt{5})/2 = 1.61803398875...$ is the *golden ratio* [6]. The approximate formula, $A(n) \sim 2.55^n$, obtained from the numerical computations up to n = 16 in [3] and n = 22 in [A048211] is consistent with the analytical results presented above. The asymptotic formula for G(n) strictly fixes the upper bound of A(n). Table-1 lists the values of A(n) obtained computationally [3, A048211]. From Table-1, we note that the ratios A(n + 1)/A(n) approach 2.533 for n = 21. This is not very different from the analytically obtained value 2.618. Due to computational limitations, it has not been possible to go beyond n = 22 [3, A048211]. It is to be noted that the ratios do not converge smoothly; they increase and decrease. This is not very surprising since, such a non-smooth behaviour is not uncommon in the world of integer sequences. Even in the case of the well-understood and celebrated Fibonacci sequence, the ratios do not converge smoothly. However, the non-smooth behaviour in the case of the Fibonacci has been completely quantified (Appendix-D has the relevant details).

When using G(n) for A(n), there is a certain amount of over counting. Farey sequence is the most exhaustive set of fractions, so it is sure to contain some terms absent in the actual circuit configurations. For large n such errors are expected to be small and ignorable. With a larger number of resistors there will be a larger number of configurations producing an ever larger set of equivalent resistances. So, the asymptotic relation derived above may be attainable for reasonably large n.

When we go beyond the series and parallel configurations (such as the bridge circuits), the Farey scheme is still applicable in providing a strict upper bound. Individual bridge circuits do not necessarily abide by the reciprocal theorem. A counterexample using 11 unit resistors is presented below.

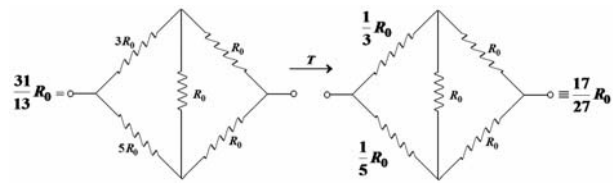

**Fig.-4. Bridge circuit violating the reciprocal theorem**

The equivalent resistance $(31/13)R_0$ after reciprocation changes to $(17/27)R_0$. The symbol $\xrightarrow{T}$ in Fig.-4., stands for reciprocation (details available in Appendix-B, containing the proof of the reciprocal theorem). We recall that for 11 resistors, a and b are bounded by Fib(12) = 144. The bridge circuits can be analyzed by the converting the delta ($\Delta$) to wye ($Y$). The three of the five bridge resistances forming the delta (say $R_1$, $R_2$, and $R_3$) get converted to the three branches of wye with the values $R_1R_2/(R_1 + R_2 + R_3)$, $R_2R_3/(R_1 + R_2 + R_3)$ and $R_3R_1/(R_1 + R_2 + R_3)$ respectively [2]. Given ($l$ + m +n) equal resistors in the delta, such converted forms in the wye produce numerators/denominators less than Fib($l$ + m + n + 1). This is so since, Fib($l$ + m + n + 1) is greater than each of the terms, $l$m, mn, n$l$ and ($l$ + m + n) respectively. The resulting wye is then combined with the remaining branches of the bridge using the usual series and parallel respectively. The delta to wye conversion complies with the bound theorem; so, any circuit containing bridges will comply with the bound theorem. As noted in the counter example in Fig.-4, individual bridge circuits can violate the reciprocal theorem. But set theoretically the bridge circuits do appear to respect the reciprocal theorem; this is evident for the bridge circuits up to n = 8. What is important is that the bridge circuits respect the bound theorem. Having established the bound theorem, the Farey scheme becomes applicable to the bridge circuits (even in the absence of the reciprocal theorem). Hence, all equivalent resistances of configurations containing bridge circuits belong to the grand set G(n). The set B(n) containing bridge circuits (in addition to the configurations produced by series and/or parallel; the set A(n) is completely contained in B(n)) has the strict bounds

$$A(n) < B(n) < G(n) = 2Farey\big(Fib(n+1);I\big) - 1.$$

The bridges may not produce too many new elements. In the case of set A(n) by the virtue of the reciprocal theorem we concluded that there are equal number of configurations on either side of 1 with the reciprocal relation for each pair. In absence of the reciprocal theorem, the same can not be said of set B(n).

The sets A(n) are for the restricted case of using all the n resistors. The Farey sequence framework is applicable to the scenario of relaxing the restriction to n *or less* resistors. Let C(n) denotes the total number of equivalent resistances obtained using one or more of the *n* equal resistors [A153588]. The set C(n) is the union of all the sets A(i), i = 1, 2, 3, …n. It is to be recalled that the sets A(i) can have elements which may not be present in the set A(j), where j ≠ i

$$C(n) = \bigcup_{i=1}^{n} A(i).$$

Each A(i) is obtained from Farey(Fib(i + 1)). We recall that the Farey sequence of a given order contains all the members of the Farey sequences of all lower orders. So, the set C(n) is strictly bounded by the Farey scheme and we have

$$A(n) < C(n) < G(n) = 2Farey\big(Fib(n+1);I\big) - 1.$$

From Table-1, we note that the ratios C(n + 1)/C(n) approach 2.514 for n = 15. For higher n we can expect a closer approach to 2.618. It is also to be noted that the ratios, C(n + 1)/C(n) converge smoothly (n ≥ 5), unlike the ones for A(n).

In Appendix-E, it is shown that certain set theoretic relations lead to $A(m) \subset A(m+3)$. Hence, it suffices to consider only the last three sets A(n − 2), A(n − 1) and A(n) in the union leading to C(n). For adequately large n,

$$C(n) \sim A(n).$$

This is reflected in the data in Table-1; C(16)/A(16) = 1.01, a difference of 1%. A better agreement is expected for larger n.

Circuits of equal resistors forming geometries such as polygons [9] and polyhedral structures [10] have been studied. From these studies, it is evident that the bound theorem is respected. Hence, any larger set of configurations involving such geometries (bridges and non-planar circuits) will continue to be bounded by the grand set, G(n).

**Concluding Remarks**

In this article we have addressed the question of the order of the set of equivalent resistances, A(n) of n equal resistors combined in series and in parallel analytically, a topic traditionally approached computationally. A recurrence relation was presented and it was shown that $2.41^n < A(n)$. As for the upper bound, it is fixed by G(n), the grand set, constructed using the Farey sequence with the Fibonacci numbers as its argument. The approximate formula, $A(n) \sim 2.55^n$, obtained from the computational data up to n = 22 is consistent with the analytical lower bound $2.41^n < A(n)$ and the strict upper bound, $A(n) < 2.618^n$ presented here. It is further shown that the Farey sequence approach, developed for the A(n) is applicable to configurations other than the series and/or parallel, namely the bridge circuits and non-planar circuits; bounds for such circuit configurations are also presented. The scheme also enables us to understand the case of C(n), the order of the set of equivalent resistances using n *or fewer* equal resistors. It is seen that C(n) converges to A(n) for large n. It would be worthwhile to carryout the computations beyond n = 22, and find the value of n for which we attain the number 2.61 (the two decimals of the upper bound 2.618). The computer programs using the symbolic package MATHEMATICA are available in Appendix-F and just need to be run on a faster computer. The results obtained can be shared at *The On-Line Encyclopedia of Integer Sequences* [4].

Several set theoretic relations among the sets A(n), are derived using simple arguments. From the set theoretic relations we concluded that all sets have 1 as its element and consequently are of odd order, with three exceptions, A(2) = 2, A(3) = 4 and A(5) = 22. It is shown that every set A(m) is completely contained in all the subsequent and larger sets, A(m + 3) along with the infinite and complete sequence of sets A(m + 5), A(m +6), A(m + 7), … and so on. However, the available set theoretic relations are silent about the nearest neighbour, A(m + 1); next-nearest neighbour, A(m + 2) and the near-neighbour A(m + 4). The decomposition of the set A(n) into three basic subsets derived from A(n − 1) enabled us to obtain the lower bound analytically. The set theoretic relations point to the complexity of estimating A(n) using combinatoric arguments. So, the Farey sequence approach presented here, is the only available means to estimate the sets A(n).

Every computational approached is constrained by a finite n, due to computer memory. The analytic results presented here works for all n; in fact the asymptotic relations only improve with larger n.

| n<br>A000027 | $m = Fib(n+1)$<br>A000045 | Farey (m)<br>A176499 | G(n)<br>$(1-\frac{1}{n})\frac{6\phi^2}{5\pi^2}\phi^{2n}-1$ | G(n)<br>$2Farey(Fib(n+1);1)-1$<br>A176502 | A(n)<br>A048211 | A(n+1)/A(n) | C(n)<br>A153588 | C(n+1)/C(n) | C(n)/A(n) |
|---|---|---|---|---|---|---|---|---|---|
| 1 | 1 | 2 | - | 1 | 1 | -- | 1 | -- | 1 |
| 2 | 2 | 3 | 0 | 3 | 2 | 2 | 3 | 3 | 1.5 |
| 3 | 3 | 5 | 3 | 7 | 4 | 2 | 7 | 2.333333333 | 1.75 |
| 4 | 5 | 11 | 10 | 17 | 9 | 2.25 | 15 | 2.142857143 | 1.666666667 |
| 5 | 8 | 23 | 30 | 37 | 22 | 2.444444444 | 35 | 2.333333333 | 1.590909091 |
| 6 | 13 | 59 | 84 | 99 | 53 | 2.409090909 | 77 | 2.2 | 1.452830189 |
| 7 | 21 | 141 | 229 | 243 | 131 | 2.471698113 | 179 | 2.324675325 | 1.366412214 |
| 8 | 34 | 361 | 614 | 633 | 337 | 2.572519084 | 429 | 2.396648045 | 1.272997033 |
| 9 | 55 | 941 | 1634 | 1673 | 869 | 2.578635015 | 1039 | 2.421911422 | 1.195627158 |
| 10 | 89 | 2457 | 4333 | 4425 | 2213 | 2.546605293 | 2525 | 2.430221367 | 1.140985088 |
| 11 | 144 | 6331 | 11459 | 11515 | 5691 | 2.571622232 | 6235 | 2.469306931 | 1.095589527 |
| 12 | 233 | 16619 | 30252 | 30471 | 14517 | 2.550869794 | 15463 | 2.480032077 | 1.065164979 |
| 13 | 377 | 43359 | 79757 | 80055 | 37017 | 2.549907006 | 38513 | 2.490655112 | 1.040413864 |
| 14 | 610 | 113159 | 210051 | 210157 | 93731 | 2.532106870 | 96231 | 2.498662789 | 1.026672072 |
| 15 | 987 | 296385 | 552741 | 553253 | 237465 | 2.533473451 | 241519 | 2.50978375 | 1.01707199 |
| 16 | 1597 | 775897 | 1453558 | 1454817 | 601093 | 2.531290927 | 607339 | 2.514663443 | 1.010391071 |
| 17 | 2584 | 2030103 | 3820388 | 3821369 | 1519815 | 2.528419063 | | | |
| 18 | 4181 | 5315385 | 10036637 | 10040187 | 3842575 | 2.528317591 | | | |
| 19 | 6765 | 13912615 | 26357608 | | 9720769 | 2.529753876 | | | |
| 20 | 10946 | 36421835 | 69196797 | | 24599577 | 2.530620468 | | | |
| 21 | 17711 | 95355147 | 181613601 | | 62283535 | 2.531894552 | | | |
| 22 | 28657 | 249635525 | 476551197 | | 157807915 | 2.533701965 | | | |

**Table-1.: Order of the sets of equivalent resistances and related data.**

**Appendix-A**

**Circuit Analysis**

Given a set of n resistors, $R_1$, $R_2$, $R_3$, … $R_n$, the net resistance in series is given by

$$R_{net-series} = \sum_1^n R_i = R_1 + R_2 + R_3 + \ldots R_n .$$

It is obvious that the net resistance in series combination is greater than the largest resistance among $R_i$. When connected in parallel we have

$$R_{net-parallel} = \frac{1}{\sum_1^n \frac{1}{R_i}} = \frac{1}{\frac{1}{R_1} + \frac{1}{R_2} + \frac{1}{R_3} + \ldots \frac{1}{R_n}} .$$

The net resistance in parallel is less than the smallest among $R_i$. The net resistance of an arbitrary circuit (using any conceivable combination series, parallel and others) made from the set of n resistors $R_i$ necessarily lies in between $R_{net-parallel}$ and $R_{net-series}$. In the case of n equal resistors of value $R_0$, the equivalent resistance in the series and parallel combinations is $nR_0$ and $(1/n)R_0$ respectively. All conceivable circuits have a resistance $(a/b)R_0$, with a/b belonging to the interval [1/n, n].

For two resistors, the above expression simplifies to

$$R_{net-parallel} = \frac{R_1 R_2}{R_1 + R_2} .$$

In the case of two resistors, we can symbolically write the net resistance as 'sum for series' and 'product/plus for the parallel' combinations respectively.

Any two terminal network of resistors can eventually be reduced to a single equivalent resistance by successive applications of resistances in series or resistances in parallel. A general network with an arbitrary number of terminals can not be reduced using only series and parallel combinations. We need a transform technique, which converts delta ($\Delta$) to wye ($Y$) and vice versa. The relevant conversion for solving the bridge is

$$R_1' = \frac{R_1 R_2}{R_1 + R_2 + R_3},$$

$$R_2' = \frac{R_1 R_3}{R_1 + R_2 + R_3}.$$

$$R_3' = \frac{R_2 R_3}{R_1 + R_2 + R_3}.$$

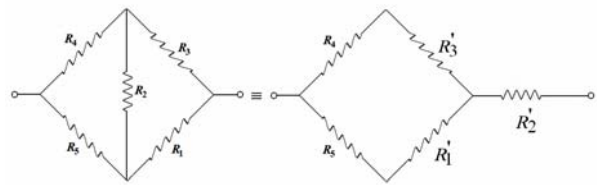

**Fig.-5. Delta to wye conversion for a bridge circuit**

The resulting wye is then combined with the remaining branches of the bridge using the usual series and parallel respectively. It is to be noted that any arbitrary network can be solved by the successive applications of series, parallel, **Δ-Y** and **Y-Δ**; no further transforms are required.

Let us now consider the sets formed by bridge circuits. An exclusive bridge circuit (EB) consists of equal resistors confined to the five arms of the bridge (the other usage is, four arms and the diagonal). The first three sets are

$$EB(5) = 1: \left\{ n = 1:1 \right\},$$

$$EB(6) = 3: \left\{ n = 6: \frac{11}{13}, 1, \frac{13}{11} \right\},$$

$$EB(7) = 17: \left\{ n = 7: \frac{2}{3}, \frac{5}{7}, \frac{3}{4}, \frac{7}{9}, \frac{16}{19}, \frac{17}{20}, \frac{19}{21}, \frac{18}{19}, 1, \frac{19}{18}, \frac{21}{19}, \frac{20}{17}, \frac{19}{16}, \frac{9}{7}, \frac{4}{3}, \frac{7}{5}, \frac{3}{2} \right\}.$$

We note EB(8) = 53 [A174285]. 1 is always an element of EB(n), n ≥ 5; since, the configurations of equal resistances in the four arms with anything in the diagonal lead to a net resistance of unity. When using at most n resistors (n or fewer resistors), we have EBS(5) = 1, EBS(6) = 3, EBS(7) = 19 and EBS(8) = 67 [A174286]. The exclusive bridge circuits can be treated as blocks and combined with other resistors in series/parallel or in bridge (and even bridge in a bridge, which requires nine or more resistors).

The sets, B(n) were obtained by including the bridge circuits in the sets A(n). Since, a minimum of five resistors are required to construct the bridges, the first four terms are unaffected. The net resistance of the five resistor bridge is 1; consequently, B(5) = 23. B(6) = 57, with the new elements; B(6) − A(6) = {1/2, 11/13, 13/11, 2}. Finally, B(7) = 151 and B(8) = 409 [A174283]. Still larger sets D(n), are obtained by including the bridge circuits in the sets C(n). The first four terms of C(n) can not be affected, as the bridge requires a minimum of five resistors. The net resistance of the five resistor bridge is 1, which is already present in C(5). Hence, D(5) = C(5) = 35. Higher sets are affected and we have, D(6) = 79, D(7) = 193 and D(8) = 489 [A174284].

**Appendix-B**

**Reciprocal Theorem**

*Theorem-1* (*Reciprocal Theorem*):
In any circuit constructed from n equal resistors (of value $R_0$) in series and/or parallel combination has an equivalent resistance $(a/b)R_0$, then the configuration obtained by changing all series connection to parallel and parallel connections to series respectively, results in a configuration, whose equivalent resistance is $(b/a)R_0$.

The proof of the reciprocal theorem is by induction similar to the one in [3]. The notation S and P stand for series and parallel respectively. The two basic blocks consisting of n resistors in series denoted by (nS) has equivalent resistance n and m resistors in parallel denoted by (mP) has equivalent resistance 1/m. Since, we are dealing with equal resistors the unit resistance $R_0$ shall not be explicitly mentioned. The transform, $T \equiv T(S \rightarrow P; P \rightarrow S)$ changes all series connections to parallel and all parallel connections to series.

The action of the transform on the two basic blocks is

$$(n_1 S) \equiv n_1,$$

$$(n_1 S) \xrightarrow{\quad T \quad} (n_1 P) \equiv \frac{1}{n_1},$$

$$(n_2 P) \equiv \frac{1}{n_2},$$

$$(n_2 P) \xrightarrow{\quad T \quad} (n_2 S) \equiv n_2.$$

The theorem is valid for the single basic blocks. Let us consider circuits made from the two basic blocks, which can be combined either in series and/or in parallel respectively, followed by the action of the transform, T

$$(n_1 S) S (n_2 P) \equiv n_1 + \frac{1}{n_2} = \frac{1 + n_1 n_2}{n_2},$$

$$(n_1 S) S (n_2 P) \xrightarrow{\quad T \quad} (n_1 P) P (n_2 S) \equiv \frac{1}{n_1} P n_2 \equiv \frac{n_2}{1 + n_1 n_2},$$

$$(n_1 S) P (n_2 P) \equiv n_1 P \frac{1}{n_2} \equiv \frac{n_1}{1 + n_1 n_2},$$

$$(n_1 S)P(n_2 P) \xrightarrow{\ T\ } (n_1 P)S(n_2 S) \equiv \frac{1}{n_1} S n_2 \equiv \frac{1 + n_1 n_2}{n_1}.$$

The equivalent resistance under the transform changes to the reciprocal of the original; i.e., $(a/b)R_0$ transforms to $(b/a) R_0$. The theorem is found to be valid for combinations of two blocks. Next we consider the case of three blocks

$$[(n_1 S)S(n_2 P)]S(n_3 P) \equiv \frac{1 + n_1 n_2}{n_2} S \frac{1}{n_3} \equiv \frac{n_2 + n_3 + n_1 n_2 n_3}{n_2 n_3},$$

$$[(n_1 S)S(n_2 P)]S(n_3 P) \xrightarrow{\ T\ } [(n_1 P)P(n_2 S)]P(n_3 S) \equiv \frac{n_2}{1 + n_1 n_2} P n_3 \equiv \frac{n_2 n_3}{n_2 + n_3 + n_1 n_2 n_3},$$

$$[(n_1 S)S(n_2 P)]P(n_3 P) \equiv \frac{1 + n_1 n_2}{n_2} P \frac{1}{n_3} \equiv \frac{1 + n_1 n_2}{n_2 + n_3 + n_1 n_2 n_3},$$

$$[(n_1 S)S(n_2 P)]P(n_3 P) \xrightarrow{\ T\ } [(n_1 P)P(n_2 S)]S(n_3 S) \equiv \frac{n_2}{1 + n_1 n_2} S n_3 \equiv \frac{n_2 + n_3 + n_1 n_2 n_3}{1 + n_1 n_2}.$$

The theorem is found to be valid for three blocks. We assume the theorem to be valid for two sub-circuits constructed from arbitrary number of m and n resistors, with equivalent resistance $(a/b) R_0$ and $(c/d) R_0$ respectively. Under the action of the transform we have

$$\left[ m : \frac{a}{b} \right] S \left[ n : \frac{c}{d} \right] \equiv \frac{a}{b} + \frac{c}{d} = \frac{ad + bc}{bd},$$

$$\left[ m : \frac{a}{b} \right] S \left[ n : \frac{c}{d} \right] \xrightarrow{\ T\ } \left[ m : \frac{b}{a} \right] P \left[ n : \frac{d}{c} \right] \equiv \frac{b}{a} P \frac{d}{c} = \frac{bd}{ad + bc},$$

$$\left[ m : \frac{a}{b} \right] P \left[ n : \frac{c}{d} \right] \equiv \frac{a}{b} P \frac{c}{d} = \frac{ac}{ad + bc},$$

$$\left[ m : \frac{a}{b} \right] P \left[ n : \frac{c}{d} \right] \xrightarrow{\ T\ } \left[ m : \frac{b}{a} \right] S \left[ n : \frac{d}{c} \right] \equiv \frac{b}{a} S \frac{d}{c} = \frac{ad + bc}{ac}.$$

Thus by induction, the reciprocal theorem has been proved.

**Appendix-C**

**Largest Numerator-Denominator Theorem**

*Theorem-2* (*Bound of a and b Theorem*):
Any circuit constructed from n equal resistors (of value $R_0$) in series and/or parallel combination has an equivalent resistance $(a/b)R_0$, such that the largest possible values of a and b are bounded by the Fib(n + 1), the $(n + 1)^{th}$ term in the Fibonacci sequence. This largest value Fib(n + 1), is obtained by taking the combinations in series and parallel incrementally.

The above theorem is mentioned in passing but without any proof [A048211]. Before proving the theorem let us see how the Fibonacci numbers, Fib(n + 1) arise [6]. Let $(1S1P1)_n$, denote a circuit constructed by connecting n resistors in series and parallel incrementally.

2 Resistors: $(1S1P1)_2 \equiv 1S1 = 1 + 1 = 2$,

3 Resistors: $(1S1P1)_3 \equiv 1S1P1 = 2P1 = \dfrac{2}{3}$,

4 Resistors: $(1S1P1)_4 \equiv 1S1P1S1 \equiv \dfrac{2}{3}S1 = \dfrac{5}{3}$.

The corresponding resistances are found to be ratios of two consecutive Fibonacci numbers [A000045]. By induction one can conclude that for n resistors we have

n Resistors: $(1S1P1)_n \equiv \begin{cases} \dfrac{Fib(n)}{Fib(n+1)} \, odd-n \\ \dfrac{Fib(n+1)}{Fib(n)} \, even-n. \end{cases}$

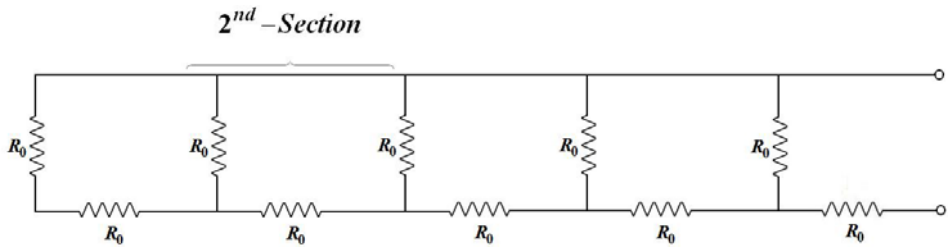

**Fig.-6. Ladder Network**

$(SP)_n$ represents a ladder network [6]. Traditionally the ladder network has n sections with two resistors in each with an equivalent resistance of Fib(2n + 1)/Fib(2n).

The ladder network starts with two resistors in series. In principle, there can be more than two resistors in series in the beginning. With two resistors in series in the beginning, we have the chain $(2SP)_n$

4 Resistors: $(2S1P1)_3 \equiv 2S1P1 = 3P1 = \dfrac{3}{4}$,

5 Resistors: $(2S1P1)_4 \equiv \dfrac{3}{4}S1 = \dfrac{7}{4}$,

n Resistors: $(2S1P1)_{n>4} \equiv \begin{cases} \dfrac{L(n-1)}{L(n-2)} \, odd - n \\[2mm] \dfrac{L(n-2)}{L(n-1)} \, even - n. \end{cases}$

L(n) are the Lucas numbers [A000204]. We consumed n resistors to reach the $(n-1)^{th}$ Lucas number, L(n − 1). The same number of resistors in the usual ladder network lead to the Fibonacci number, Fib(n + 1). We note, that L(n − 1) < Fib(n + 1). Likewise, any other deviation in the beginning of the ladder network produces numerators/denominators less than the Fib(n + 1). This is also true for the deviations in the middle and at the end of the ladder network, though the details do vary.

In the proof of the reciprocal theorem we noticed that circuits assembled from two and three basic blocks results in numerators/denominators such as $(1 + n_1 n_2)$ and $(n_2 + n_3 + n_1 n_2 n_3)$ for $n = (n_1 + n_2)$ and $n = (n_1 + n_2 + n_3)$ respectively. Such numerators/denominators are less than Fib(n + 1). So, circuits made from blocks have smaller numerators/denominators.

We note, that the variants of the ladder network and circuits made from blocks produce equivalent resistances a/b such that a and b are both less than the Fibonacci(n + 1). This is the heuristic proof of the reciprocal theorem.

**Appendix-D**

**Mathematical Notes**

We shall briefly state the relevant properties of the Fibonacci numbers and the Haros-Farey sequence.

**Fibonacci Numbers**

Fibonacci numbers are the sequence of numbers, 1, 1, 2, 3, 5, 8, 13, 21, 34, 55, … [A000045], for n = 1, 2, 3, 4, 5, … [A000027]. Successive Fibonacci numbers are obtained by taking the sum of two preceding numbers, which is expressed as the linear recurrence relation

$$F_n = F_{n-1} + F_{n-2},$$

for n > 3, with $F_1 = F_2 = 1$.

The linear recurrence relations are solved by introducing a constant ratio, $\lambda = F_n / F_{n-1}$, between any two successive terms, for adequately large n. This leads to the quadratic equation $\lambda^2 = \lambda + 1$, with one of the roots as $\phi = (1 + \sqrt{5})/2$. The ratios of the pair of Fibonacci numbers, Fib(n + 1)/Fib(n) rapidly converges to the golden ratio, $\phi = (1 + \sqrt{5})/2 = 1.61803398874....$ This ratio occurs in diverse situations and hence has been named as the *golden ratio* or even the *divine proportion* [6]. Its value to 61 decimal places is

1.6180339887498948482045868343656381177203091798805762862135448.

Even for n as small as 11 (corresponding to 10 equal resistors in the present study) the ratio agrees with the first three decimal places. From Table-2, it is to be noted that the ratios do not converge smoothly; they increase and decrease alternately. This can be seen by taking the difference of any two successive ratios and simplifying the expression using the Cassini's identity (which is a special case of the Catalans identity[11]), we have

$$\frac{F_{n+1}}{F_n} - \frac{F_n}{F_{n-1}} = \frac{F_{n+1}F_{n-1} - F_n^2}{F_n F_{n-1}} = \frac{(-1)^n}{F_n F_{n-1}}.$$

The above expression points to the change in sign leading to the non-smooth behaviour and also provides the magnitude of this difference.

Computation of an arbitrary Fibonacci number is facilitated by the closed form expression

$$F_n = \left[ \frac{\phi^n}{\sqrt{5}} \right],$$

where […] is the nearest integer function. This remarkable result is a consequence of the linear recurrence relation satisfied by the Fibonacci numbers, and works for all natural numbers, n.

| n\n\nA000027 | Fibonacci (n)\n\nA000045 | $\frac{\phi^n}{\sqrt{5}}$ | Fib(n+1)/Fib(n) |
|---|---|---|---|
| 1 | 1 | 0.723606798 | - |
| 2 | 1 | 1.170820393 | 1 |
| 3 | 2 | 1.894427191 | 2 |
| 4 | 3 | 3.065247584 | 1.5 |
| 5 | 5 | 4.959674775 | 1.666666667 |
| 6 | 8 | 8.024922359 | 1.6 |
| 7 | 13 | 12.98459713 | 1.625 |
| 8 | 21 | 21.00951949 | 1.615384615 |
| 9 | 34 | 33.99411663 | 1.619047619 |
| 10 | 55 | 55.00363612 | 1.617647059 |
| 11 | 89 | 88.99775275 | 1.618181818 |
| 12 | 144 | 144.0013889 | 1.617977528 |
| 13 | 233 | 232.9991416 | 1.618055556 |
| 14 | 377 | 377.0005305 | 1.618025751 |
| 15 | 610 | 609.9996721 | 1.618037135 |
| 16 | 987 | 987.0002026 | 1.618032787 |
| 17 | 1597 | 1596.999875 | 1.618034448 |
| 18 | 2584 | 2584.000077 | 1.618033813 |
| 19 | 4181 | 4180.999952 | 1.618034056 |
| 20 | 6765 | 6765.00003 | 1.618033963 |
| 21 | 10946 | 10945.99998 | 1.618033999 |
| 22 | 17711 | 17711.00001 | 1.618033985 |
| 23 | 28656 | 28656.99999 | 1.617977528 |
| 24 | 46368 | 46368 | 1.618090452 |
| 25 | 75025 | 75025 | 1.618033989 |
| 26 | 121393 | 121393 | 1.618033989 |
| 27 | 196418 | 196418 | 1.618033989 |
| 28 | 317811 | 317811 | 1.618033989 |
| 29 | 514229 | 514229 | 1.618033989 |
| 30 | 832040 | 832040 | 1.618033989 |
| 31 | 1346269 | 1346269 | 1.618033989 |

**Table-2.: Fibonacci numbers and their ratios and the closed form formula.**

A closely related set of numbers (again following the same recurrence relation) are the Lucas numbers, 1, 3, 4, 7, 11, 18, 29, 47, 76, 123, …[ A000204].  The closed form expression for the Lucas numbers is

$$L_n = \left[\phi^n\right].$$

The common feature of all such number sequences is the golden ratio (independent of the seeds) and the closed form expressions (which do depend on the choice of seeds). One can take any two numbers, a and b as seeds and generate the corresponding sequence of Fibonacci-type numbers (called as the generalized Fibonacci numbers [11]) with the formula

$$f_n = \frac{1}{2}\left[(3a - b)F_n + (b - a)L_n\right].$$

The lore surrounding the Fibonacci numbers is gigantic and there is even a journal, *The Fibonacci Quarterly,* devoted to the study of integers with special properties [12].

**Haros-Farey Sequence**

The Farey sequence of order m (a natural number) is the set of irreducible rational numbers a/b with $0 \le a \le b \le m$.  The fractions are traditionally arranged in order of increasing size [7, 8, A005728].  The first few are

$$Farey(1) = 2: \left\{ m = 1 : \frac{0}{1}, \frac{1}{1} \right\},$$

$$Farey(2) = 3: \left\{ m = 2 : \frac{0}{1}, \frac{1}{2}, \frac{1}{1} \right\},$$

$$Farey(3) = 5: \left\{ m = 3 : \frac{0}{1}, \frac{1}{3}, \frac{1}{2}, \frac{2}{3}, \frac{1}{1} \right\},$$

$$Farey(4) = 7: \left\{ m = 4 : \frac{0}{1}, \frac{1}{4}, \frac{1}{3}, \frac{1}{2}, \frac{2}{3}, \frac{3}{4}, \frac{1}{1} \right\},$$

$$Farey(5) = 11: \left\{ m = 5 : \frac{0}{1}, \frac{1}{5}, \frac{1}{4}, \frac{1}{3}, \frac{2}{5}, \frac{1}{2}, \frac{3}{5}, \frac{2}{3}, \frac{3}{4}, \frac{4}{5}, \frac{1}{1} \right\}.$$

Each Farey sequence starts with zero (denoted by 0/1) and ends with 1 (denoted by 1/1), and contains all conceivable fractions whose denominators are less than or equal to m.  The Farey sequence of order m contains all of the members of the Farey sequences of all lower orders.  In particular, Farey(m) contains all the members of Farey(m − 1), and also contains an additional fraction for each number that is less than m and coprime to m. Thus Farey(6) consists of Farey(5) together with the fractions 1/6 and 5/6. Except for Farey(1), each set Farey(m) contains odd number of elements and the middle term is always 1/2. The sequence is generally attributed to Farey (1816), but an earlier publication can be traced to Haros (1802) (see [8] for comprehensive historical and technical accounts). Hence, the name Haros-Farey sequence.

Unlike, the Fibonacci numbers there is no known closed form expression for the Farey sequence. However, the asymptotic limit for the Farey(m) is obtained using the Euler's totient function, $\varphi(i)$

$$Farey(m) = 1 + \sum_{i=1}^{m} \varphi(i) \sim \frac{3}{\pi^2} m^2 + O(m \log m).$$

The above asymptotic formula (even without the logarithmic term) differs by less than 1% for m as small as 144 = Fibonacci(12), corresponding to 11 resistors, in the present study. For higher m the agreement is still better. In our study, we shall omit the logarithmic correction and replace the similarity symbol ~ with the equality symbol.

It is generally accepted that the Farey sequence is uniformly distributed in [0, 1]; recall that 1/2 is always the middle term [7, 8]. Let I = [α, β] be a subinterval of [0, 1], the Farey fractions of order m from I is related to the length of the subinterval, $|I| = |\beta - \alpha| = (\beta - \alpha)$ by the relation

$$Farey(m; I) = |I| \times Farey(m) = 3(\beta - \alpha) \frac{1}{\pi^2} m^2.$$

There is a connection between the Farey sequence and the Riemann hypothesis [8]. There are two equivalent formulations of the Riemann hypothesis based on the distribution of the Farey(n). It would be worthwhile to study the distribution of A(n) and check for possible correlations between A(n) and Farey(n). Any deeper connection (if any) between the A(n) with Farey may relate A(n) to the Riemann hypothesis.

**The Grand Set**
The Grand Set, G(n) by construction is the union of the set Farey(Fib(n + 1); I) and its reciprocal, where the interval, I = [1/n, 1]; consequently, G(n) contains all the rational numbers in the interval [1/n, n], whose numerators and denominators are bounded by Fibonacci(n + 1).

$$G(n) = \left\{ Farey\big(Fib(n+1); I\big) \right\} \bigcup \left\{ 1 / Farey\big(Fib(n+1); I\big) \right\}$$

The order of the Grand Set is $G(n) = 2 Farey\big(Fib(n+1); I\big) - 1$. The first few sets are:

$GrandSet(1) = 1: \{n = 1:1\},$

$GrandSet(2) = 3: \left\{n = 2: \dfrac{1}{2}, 1, 2\right\},$

$GrandSet(3) = 7: \left\{n = 3: \dfrac{1}{3}, \dfrac{1}{2}, \dfrac{2}{3}, 1, \dfrac{3}{2}, 2, 3\right\},$

$GrandSet(4) = 17: \left\{n = 4: \dfrac{1}{4}, \dfrac{1}{3}, \dfrac{2}{5}, \dfrac{1}{2}, \dfrac{3}{5}, \dfrac{2}{3}, \dfrac{3}{4}, \dfrac{4}{5}, 1, \dfrac{5}{4}, \dfrac{4}{3}, \dfrac{3}{2}, \dfrac{5}{3}, 2, \dfrac{5}{2}, 3, 4\right\}.$

The Grand set grows rapidly, 1, 3, 7, 17, 37, 99, 243, 633, 1673, 4425, 11515, 30471, 80055, 210157, 553253, 1454817, … [A176502], and its computation is constrained by the computer memory. In such a situation, one can use the approximate expressions or the asymptotic relation. The comparison of the same is presented in the Table-3.

| n A000027 | $m = Fib(n+1)$ A000045 | $Farey(m)$ A176499 | $Farey(m; I)$ A176501 | Grand Set | | |
|---|---|---|---|---|---|---|
| | | | | $2 Farey(m) - 3$ A176500 | $2 Farey(m; I) - 1$ A176502 | $(1 - \dfrac{1}{n}) \dfrac{6}{\pi^2} \left(\dfrac{\phi^{n+1}}{\sqrt{5}}\right)^2 - 1$ Rounded |
| 1 | 1 | 2 | 1 | 1 | 1 | - |
| 2 | 2 | 3 | 2 | 3 | 3 | 0 |
| 3 | 3 | 5 | 4 | 7 | 7 | 3 |
| 4 | 5 | 11 | 9 | 19 | 17 | 10 |
| 5 | 8 | 23 | 19 | 43 | 37 | 30 |
| 6 | 13 | 59 | 50 | 115 | 99 | 84 |
| 7 | 21 | 141 | 122 | 279 | 243 | 229 |
| 8 | 34 | 361 | 317 | 719 | 633 | 614 |
| 9 | 55 | 941 | 837 | 1879 | 1673 | 1634 |
| 10 | 89 | 2457 | 2213 | 4911 | 4425 | 4333 |
| 11 | 144 | 6331 | 5758 | 12659 | 11515 | 11459 |
| 12 | 233 | 16619 | 15236 | 33235 | 30471 | 30252 |
| 13 | 377 | 43359 | 40028 | 86715 | 80055 | 79757 |
| 14 | 610 | 113159 | 105079 | 226315 | 210157 | 210051 |
| 15 | 987 | 296385 | 276627 | 592767 | 553253 | 552741 |
| 16 | 1597 | 775897 | 727409 | 1551791 | 1454817 | 1453558 |
| 17 | 2584 | 2030103 | 1910685 | 4060203 | 3821369 | 3820388 |
| 18 | 4181 | 5315385 | 5020094 | 10630767 | 10040187 | 10036637 |
| 19 | 6765 | 13912615 | | 27825227 | | 26357608 |
| 20 | 10946 | 36421835 | | 72843667 | | 69196797 |
| 21 | 17711 | 95355147 | | 190710291 | | 181613601 |
| 22 | 28657 | 249635525 | | 499271047 | | 476551197 |

**Table-3.: The Grand Set.**

**Appendix-E**

**Some Set Theoretic Properties of A(n)**

We shall note and derive some properties of the sets A(n).

*Scaling Property*:
If a/b is a member of A(m), then we can construct the resistances k(a/b) and (1/k)(a/b) using k such blocks in series and parallel respectively, using km number of unit resistors

$$kA(m) \in A(km) \quad and \quad \frac{1}{k}A(m) \in A(km).$$

A block of i resistors in series has an equivalent resistance i. If i such blocks are combined in parallel we get back the unit resistance. From this we conclude that

$$1 \in A(i^2).$$

The above result can equivalently be obtained by taking i blocks in series, each containing i unit resistors in parallel. Once the unit resistor has been obtained, using $i^2$ resistors (or much less as we shall soon see), we can use it to construct other equivalent resistances. Every set A(m) is made from m unit resistors. The same set can be replicated by using m-number of unit resistors constructed with $i^2$ resistors. So,

$$A(m) \subset A(i^2 m).$$

Whenever 1 belongs to some set A(i), we label it as $1_i$, to indicate that it has been constructed from i number of basic unit resistors, $R_0$.

*Translation Property*:
Using $1_i$, we have the translation property, $1 \in A(i) \Rightarrow 1 \in A(i+3)$. This can be seen by taking either of the following combination of $1_i$ with 3 basic unit resistors

$$(1S1)P(1S1_i) \equiv 2P2 = \frac{2 \times 2}{2+2} = \frac{4}{4} = 1,$$

$$(1P1)S(1P1_i) \equiv \left(\frac{1}{2}\right)S\left(\frac{1}{2}\right) = \frac{1}{2} + \frac{1}{2} = 1.$$

So whenever, $1 \in A(i)$, it follows that $1 \in A(i+3)$. We shall use this translation property to prove the theorem, that 1 belongs to all A(n) barring three exceptions.

*Theorem*:

$$1 \in A(n), n \neq 2, n \neq 3, n \neq 5 .$$

From an exhaustive search (or otherwise) we know that 1 belongs to A(6), A(7) and A(8). Using the translational property, 1 also belongs to A(9), A(10) and A(11); A(12), A(13) and A(14); and so on. Thus we conclude that 1 belongs to all A(n), n ≥ 6. As for the lower A(i), 1 belongs to A(1) and A(4); and 1 does not belong to A(2), A(3) and A(5). Hence the theorem is proved.

In passing we note that, an exhaustive search is not always required. Two resistors in parallel lead to 1/2; and two such blocks in series lead to 1 and hence, $1 \in A(4)$. The set A(3) contains 1/3 and 2/3; combining these two blocks in series gives 1, implying $1 \in A(6)$. Combining the 1/2 present in A(2) with the 1/2 in A(5) in series, we conclude that $1 \in A(7)$. Similarly 1/4 and 3/4 are present in A(4) and lead to $1 \in A(8)$.

Recalling that all elements in A(n) have a reciprocal pair (a/b with b/a) and 1 is its own partner; presence of element 1 implies that A(n) is always odd with the exception of A(2) = 2, A(3) = 4 and A(5) = 22. We recall that, except for Farey(1), each set Farey(m) contains odd number of elements.

*Corollary*:

$$\frac{1}{2} \in A(n), n \neq 1, n \neq 3, n \neq 4, n \neq 6 .$$

The parallel combination of 1 basic unit resistor, $R_0$ with $1_i$ (i = 4 and i ≥ 6) results in an equivalent resistance of 1/2, (since, $1 P 1_i \equiv (1 \times 1_i)/(1 + 1_i) = 1/2$), which implies that $1/2 \in A(i+1), i = 4, i \geq 6$. The corollary is proved for n = 5 and all n ≥ 7. Resorting to the exhaustive search, we note that 1/2 belongs to A(2); the four exceptional sets are A(1), A(3), A(4) and A(6), which do not contain the element 1/2. We recall that, except for Farey(1), each set Farey(m) contains 1/2.

We examined the occurrence of 1 in A(i), since 1 is the basic unit and all other resistances can be constructed from it. The element 1/2 was examined in the spirit of Farey sequence. Remaining elements (infinite in number) shall be discussed collectively using the modular property.

We constructed $1_i$ from i basic unit resistors (i = 4, and i ≥ 6). Any set A(m) can be constructed from m number of $1_i$, using mi number of resistors, consequently

$$A(m) \subset A(mi), i = 1, i = 4, i \geq 6 .$$

The above statement is silent about i = 2, 3 and 5. The argument, (m i) is multiplicative, giving no information about the near or immediate neighbours of A(m).

Additive statements have the arguments of the type (m + i) and when they exist, they provide information about the neighbours of A(m). In the present context the additive statements are more informative and override the multiplicative statements.

*Modular Theorem*:

$$A(m) \subset A(m+3)$$
$$A(m) \subset A(m+i), \quad i \geq 5.$$

Every set A(m) is constructed from m basic unit resistors $R_0$. Let us replace any one of these basic unit resistors with $1_i$ (i = 4 and i ≥ 6). We are thus reproducing the complete set A(m) using a larger number of resistors, which is (m + i − 1). Consequently, $A(m) \subset A(m+i-1), i = 4, i \geq 6$. Thus the modular theorem is proved stating that every set A(m) is completely contained in all the subsequent and larger sets, A(m + 3) along with the infinite and complete sequence of sets A(m + 5), A(m +6), A(m + 7), … and so on. However, it is very curious to note that the infinite range theorem is silent about the three important sets: the nearest neighbour, A(m + 1); next-nearest neighbour, A(m + 2) and the near-neighbour A(m + 4).

From the modular relation, $A(m) \subset A(m+i), i \geq 5$, we conclude that

$$A(n-5) \subset A(n) \cap A(n+1), n \geq 6.$$

This is the closest we can get to know the overlap between A(n) and its nearest neighbour A(n + 1). The modular relations appear to point to a five-term recurrence relation!

One immediate consequence of the modular theorem is on the sets C(n), obtained by taking the union of A(i)

$$C(n) = \bigcup_{i=1}^{n} A(i) \equiv \bigcup_{i=n-2}^{n} A(i) \equiv A(n-2) \cup A(n-1) \cup A(n).$$

It suffices to consider only the last three sets A(n − 2), A(n − 1) and A(n) in the union. Hence it is not surprising that the ratios C(n)/A(n) are close to 1.

*Decomposition of* A(n):
The set A(n) can be constructed by adding the $n^{th}$ resistor to the set A(n − 1 ). This addition can be done in three distinct ways and results in three basic subsets of A(n). Treating the elements of A(n − 1 ) as single blocks, the $n^{th}$ resistor can be added either in series or in parallel. We call these two sets as series set and parallel set and denoted them by 1SA(n − 1) and 1PA(n − 1) respectively. The $n^{th}$ resistor can also be added somewhere within the A(n − 1) blocks, and we call this set as the cross set and denote it by $1 \otimes A(n)$. The set A(n) is the union of the three sets formed by different ways of adding the $n^{th}$ resistor

$$A(n) \equiv 1PA(n-1) \cup 1SA(n-1) \cup 1 \otimes A(n-1).$$

This decomposition is very illustrative, and enables us to understand some of the properties of A(n). All the elements of the parallel set are strictly less than 1 (since 1P(a/b) = a/(a + b) < 1) and that of the series set are strictly greater than 1 (since 1S(a/b) = (a + b)/b > 1). So,

$$1PA(n-1) \cap 1SA(n-1) \equiv \varnothing.$$

The series and the parallel sets each have exactly A(n − 1) number of configurations and the same number of equivalent resistances. Let c/d and d/c be any reciprocal pair (ensured by the reciprocal theorem) in A(n − 1), then it is seen that 1P(c/d) = c/(c + d) and 1P(d/c) = d/(c + d) belong to the set 1P A(n − 1); and 1S(c/d) = (c + d)/d and 1S(d/c) = (c + d)/c belong to the set 1S A(n − 1). This shows that all the reciprocal partners of 1PA(n − 1) always belong to 1SA(n − 1) and vice versa. These two disjoint sets contribute 2A(n − 1) number of elements to A(n) and are the source of $2^n$. The order of the cross set, $1 \otimes A(n)$ is A(n + 1) − 2A(n) and results in the sequence, 0, 0, 0, 1, 4, 9, 25, 75, 195, 475, 1265, 3135, ... [A176497]. It is the cross set which takes the count beyond $2^n$ to $2.53^n$ numerically and maximally to $2.61^n$, strictly fixed by the Farey scheme. For n ≥ 7, all the three basic sets have odd number of elements, since A(n) is odd for n ≥ 6.

*Complementary Property*: Every element in any A(n) has a reciprocal pairs (1 being its own partner). Every set A(n), n ≥ 3, also has some complementary pairs such that their sum is equal to 1. For example 1 pair (1/3, 2/3) in A(3); two pairs (1/4, 3/4) and (2/5, 3/5) in A(4). The element 1/2 (n ≠ 1, n ≠ 3, n ≠ 4 and n ≠ 6) can be treated as its own complementary partner. We shall soon conclude that each element of the set 1PA(n − 1) has a complementary partner in 1PA(n − 1) itself. The element 1/2 is the only exception and may be treated as its own partner. By the reciprocal theorem, the elements c/d and d/c occur as reciprocal pairs in every A(n − 1); then in 1P A(n − 1) we have

$$\left(1P\frac{c}{d}\right) + \left(1P\frac{d}{c}\right) = \frac{c}{c+d} + \frac{d}{c+d} = 1.$$

Thus all the elements (except the element 1/2) of 1PA(n − 1) have a complementary partner in 1PA(n − 1) itself. For n ≥7, the number of such pairs is (A(n − 1) − 1)/2, since A(n − 1) is odd for n ≥7 and $1/2 \in A(n), n \geq 7$. It is obvious that the set 1SA(n − 1) does not have complementary pairs. It is to be recalled that elements in the Farey set are complementary with respect 1; the median point 1/2 is the only exception, and may be treated as its own partner.

The cross set is not straightforward, as it is generated by placing the $n^{th}$ resistor anywhere within the blocks of A(n − 1). It is the source of all the extra configurations, which do not result in any new equivalent resistances.  For, for n > 6, the cross set has at least A(n − 2) elements, since A(n − 1) has A(n − 2) connections corresponding to $1 \otimes A(n - 2)$.  By decomposition, we note that the element 1 can belong only to the cross set and not the other two (since all elements of 1PA(n − 1) are less than 1 and all the elements of 1SA(n − 1) are greater than 1)

$$1 \notin 1PA(n-1),$$
$$1 \notin 1SA(n-1),$$
$$1 \in 1 \otimes A(n-1), n \geq 6.$$

We noted that the two disjoint sets, 1PA(n − 1) and 1SA(n − 1) are reciprocal to each other.  Consequently all elements in $1 \otimes A(n - 1)$, have their reciprocal partners in $1 \otimes A(n - 1)$ itself; 1 is its own partner.  The cross set is expected to be dense near 1 with few of its elements below half (recall that 1/2 is contained in 1PA(n), $n \geq 6$, and not a member of the cross sets).  This is reflected in the fact that cross sets up to $1 \otimes A(7)$ do not have a single element below half.  The successive cross sets have, 1, 6, 9, 24, 58, 124, 312 elements respectively [A176498], a small percentage compared to the size of the cross set [A176497].

It is straightforward to carry over the set theoretic relations to the bridge circuits sets [A174283]; since, $A(n) \subset B(n)$.  Unlike A(i) the B(i) have the additional feature $1 \in B(5)$.  So, the various statements, get accordingly modified; in particular we have

$$1 \in B(n), n \neq 2, n \neq 3,$$

$$\frac{1}{2} \in B(n), n \neq 1, n \neq 3, n \neq 4,$$

$$B(m) \subset B(mi), i = 1, i \geq 4,$$

$$B(m) \subset B(m+i), i \geq 3,$$

$$B(n-3) \subset B(n) \cap B(n+1), n \geq 4.$$

*The Octet Set*:
If a/b is a member of A(i) and c/d is a member of A(j), then we have the eight-element set $A^{i,j}$; $A^{i,j} \subset A(i+j)$, where

$$A^{i,j} \equiv \left\{ \frac{bd}{ad+bc}, \frac{bc}{ac+bd}, \frac{ad}{bd+ac}, \frac{ac}{bc+ad}, \frac{ad+bc}{bd}, \frac{ac+bd}{bc}, \frac{bd+ac}{ad}, \frac{bc+ad}{ac} \right\}.$$

The set $A^{i,j}$ contains all the eight possible elements, which can be obtained by the series/parallel combinations of a/b and c/d and their reciprocals each respectively.

**Computer Programs in MATHEMATICA**

The problem of resistor networks is intrinsically a computational problem. The following programs have been written using the symbolic package MATHEMATICA. They just need to be run on a faster computer. The results obtained can be shared at *The On-Line Encyclopedia of Integer Sequences* (OEIS) [4]; additional details about the OEIS are available in Appendix-G.

```
(* n Equal Resistors connected in Series and/or Parallel*)
NumberResistors = 4;
ClearAll[CirclePlus, CircleTimes];
SetAttributes[{CirclePlus, CircleTimes}, {Flat,Orderless}];
SeriesCircuit[a_, b_]:= a⊕b;
ParallelCircuit[a_, b_]:= a⊗b;
F[a_,b_]:= Flatten[Outer[SeriesCircuit, a, b]⋃Outer[ParallelCircuit, a, b], 2];
S={{R}, {R⊕R, R⊗R}};
Do[SX = F[S[[1]], S[[i-1]]];
    Do[SX = Flatten[SX⋃F[S[[k]], S[[i-k]]], 2]; , {k, 2, i/2}];
    S = S⋃{SX}; , {i, 3, NumberResistors}];
S[[NumberResistors]] (*This line displays the Full Set of Configurations*)
Print[StringForm["NumberResistors = ``, NumberConfigurations = ``", NumberResistors,
Dimensions[S[[NumberResistors]]]]]
SetAttributes[{CirclePlus, CircleTimes},{NumericFunction, OneIdentity}];
a_⊕b_ := a + b;
a_⊗b_ := a*b/(a+b);
CirclePlus[x_]:= x;
CircleTimes[x_]:= x;
(*Print[S[[NumberResistors]]/.R→1]*) (*This line displays the Set of Equivalent Resistances
corresponding to the Set of Configurations*)
(*Print[Union[S[[NumberResistors]]/.R→1]]*) (*This line displays the Full Set of Equivalent
Resistances*)
Print[StringForm["NumberResistors = ``, NumberConfigurations = ``, NumberEquivalentResistances =
``, CPU time in seconds = ``", NumberResistors, Dimensions[S[[NumberResistors]]],
Dimensions[Union[S[[NumberResistors]]]], TimeUsed[*]]]
{R⊕R⊗ (R⊕R), R⊕R⊗R⊗R, R⊗R⊕R⊗R, R⊕R⊕R⊕R,
 R⊗ (R⊕R⊗R), R⊗ (R⊕R⊕R), (R⊕R) ⊗ (R⊕R), R⊗R⊗ (R⊕R), R⊗R⊗R⊗R}
NumberResistors = 4, NumberConfigurations = {10},
NumberEquivalentResistances = {9}, CPU time in seconds = 0.031`
```

**Program-1.:** This Program in MATHEMATICA is designed to compute the "Set of Configurations" [A00084] and the "Set of Equivalent Resistances" [A048211] of n Equal Resistors connected in Series and/or Parallel. The output is shown for n = 4.

```
(* At most n Equal Resistors (n or fewer Resistors) connected in Series and/or Parallel*)
NumberResistors = 4;
ClearAll[CirclePlus, CircleTimes];
SetAttributes[
    {CirclePlus, CircleTimes}, {Flat, Orderless}];
SeriesCircuit[a_, b_]:= a⊕b;
ParallelCircuit[a_, b_]:= a⊗b;
F[a_,b_]:=
    Flatten[Outer[SeriesCircuit, a, b] ⋃ Outer[ParallelCircuit, a, b], 2];
S = {{R}, {R⊕R, R⊗R}};
Do[SX = F[S[[1]], S[[i-1]]];
    Do[SX = Flatten[SX ⋃ F[S[[k]], S[[i-k]]], 2]; , {k, 2, i/2}];
    S = S ⋃ {SX}; , {i, 3, NumberResistors}];
SetConfigurations = Union[Flatten[Table[S, {i, NumberResistors}]]];
Print[StringForm["NumberResistors = ``, SetConfigurations = ``, NumberConfigurations =
``", NumberResistors, SetConfigurations, Dimensions[SetConfigurations] ]] (*This line
displays the Full Set of Configurations*)
SetAttributes[{CirclePlus, CircleTimes}, {NumericFunction, OneIdentity}];
a_⊕b_ := a + b;
a_⊗b_ := a*b/(a + b);
CirclePlus[x_]:= x;
CircleTimes[x_]:= x;
SetResistances = Union[SetConfigurations/.R → 1];
Print[StringForm["NumberResistors = ``, NumberConfigurations = ``,
NumberEquivalentResistances = ``, CPU time in seconds = ``"
                , NumberResistors, Dimensions[SetConfigurations],
Dimensions[SetResistances], TimeUsed[%] ]]
NumberResistors = 4, SetConfigurations = {R, R⊕R, R⊕R⊗R, R⊕R⊗(R⊕R), R⊕R⊗R⊗R,
    R⊗R⊕R⊗R, R⊕R⊕R, R⊕R⊕R⊗R, R⊕R⊕R⊕R, R⊗R, R⊗(R⊕R), R⊗(R⊕R⊗R), R⊗(R⊕R⊕R),
    (R⊕R)⊗(R⊕R), R⊗R⊕R, R⊗R⊗(R⊕R), R⊗R⊗R⊗R}, NumberConfigurations = {17}
NumberResistors = 4, NumberConfigurations = {17},
NumberEquivalentResistances = {15}, CPU time in seconds = 0.031`
```

**Program-2.:** This Program in MATHEMATICA is designed to compute the "Set of Configurations" [A058351] and the "Set of Equivalent Resistances" [A153588] of at most n Equal Resistors (n or fewer Resistors) connected in Series and/or Parallel. The output is shown for n = 4.

```
(*Order of Haros-Farey Sequence*)
m = 5;
Fareym = 1 + Sum[EulerPhi[i], {i, 0, m}];
Print[StringForm["m = ``, Order of Farey Set = ``, CPU time in seconds = ``", m, Fareym,
TimeUsed[%] ]]
m = 5, Order of Farey Set = 11, CPU time in seconds = 7.758821796849391`*^-19
```

**Program-3.:** This Program in MATHEMATICA is designed to compute the Haros-Farey Sequence [A005728] (widely known as the Farey Sequence) using the Euler's Totient Function. The output is shown for m = 5.

```
(*Haros-Farey Sequence with its Set*)
m = 5;
FareySet = Union[Flatten[Table[a/b, {b, m}, {a, 0, b}]]];
Print[StringForm["m = ``, FareySet = ``", m, FareySet ]] (*This line displays the Farey Set*)
Print[StringForm["m = ``, Order of Farey Set = ``, CPU time in seconds = ``", m,
Dimensions[FareySet], TimeUsed[%] ]]
```

$$m = 5, \; FareySet = \left\{ 0, \frac{1}{5}, \frac{1}{4}, \frac{1}{3}, \frac{2}{5}, \frac{1}{2}, \frac{3}{5}, \frac{2}{3}, \frac{3}{4}, \frac{4}{5}, 1 \right\}$$

m = 5, Order of Farey Set = {11}, CPU time in seconds = 0.047`

**Program-4.:** This Program in MATHEMATICA is designed to compute the Haros-Farey Sequence (widely known as the Farey Sequence) [A005728], along with the complete Set. The output is shown for m = 5.

```
(* Order of the Haros-Farey Sequence whose argument is Fibonacci(n + 1); Farey(m) where m = Fibonacci (n + 1)*)
n = 5;
m = Fibonacci[n + 1];
FareyFibonacci = 1 + Sum[EulerPhi[i], {i, 0, m}];
Print[StringForm["n = ``, m = Fibonacci(n + 1) = ``, Order of Farey Fibonacci Set = ``, CPU
time in seconds = ``", n, m, FareyFibonacci, TimeUsed[%] ]]
```

n = 5, m = Fibonacci(n + 1) = 8, Order of Farey
  Fibonacci Set = 23, CPU time in seconds = 4.438452643612534`*^-19

**Program-5.:** This Program in MATHEMATICA is designed to compute the Haros-Farey Sequence whose argument is Fibonacci(n + 1); Farey(m) where m = Fibonacci (n + 1), using the using the Euler's Totient Function. Farey(Fibonacci(n + 1)) is the Sequence A176499. The output is shown for n = 5.

```
(*Haros-Farey Sequence with its Set whose argument is Fibonacci(n + 1); Farey(m) where m = Fibonacci (n + 1)*)
n = 5;
m = Fibonacci[n + 1];
FareyFibonacci = Union[Flatten[Table[a/b, {b, m}, {a, 0, b}]]];
Print[StringForm["n = ``, m = Fibonacci(n + 1) = ``, Farey Fibonacci = ``, Order of
Farey Fibonacci Set = ``", n, m, FareyFibonacci, Dimensions[FareyFibonacci] ]]
(*This line prints full sets with their orders*)
Print[StringForm["n = ``, m = Fibonacci(n + 1) = ``, Order of Farey Fibonacci Set = ``,
CPU time in seconds = ``",
                n, m, Dimensions[FareyFibonacci], TimeUsed[%] ]]
```

n = 5, m = Fibonacci(n + 1) = 8,
$$\left\{ 0, \frac{1}{8}, \frac{1}{7}, \frac{1}{6}, \frac{1}{5}, \frac{1}{4}, \frac{2}{7}, \frac{1}{3}, \frac{3}{8}, \frac{2}{5}, \frac{3}{7}, \frac{1}{2}, \frac{4}{7}, \frac{3}{5}, \frac{5}{8}, \frac{2}{3}, \frac{5}{7}, \frac{3}{4}, \frac{4}{5}, \frac{5}{6}, \frac{6}{7}, \frac{7}{8}, 1 \right\}$$
Order of Farey Fibonacci Set = {23}, CPU time in seconds = 0.016`

**Program-6.:** This Program in MATHEMATICA is designed to compute the Haros-Farey Sequence whose argument is Fibonacci(n + 1); Farey(m) where m = Fibonacci (n + 1), along with its set. Farey(Fibonacci(n + 1)) is the Sequence A176499. The output is shown for n = 5.

**Page 29 of 37**

```
(*Order of the Grand Set of rational numbers, whose numerators and denominators are bounded by Fibonacci(n + 1)*)
n = 5;
m = Fibonacci[n + 1];
FareyFibonacci = 1 + Sum[EulerPhi[i], {i, 0, m}];
Print[StringForm["n = ``, m = Fibonacci(n + 1) = ``, Order of Farey Fibonacci Set = ``,
Order of Grand Set = 2*Farey(m) - 3 = ``, CPU time in seconds = ``",
                n, m, FareyFibonacci, 2*FareyFibonacci - 3, TimeUsed[%] ]]
n = 5, m = Fibonacci(n + 1) = 8, Order of Farey Fibonacci Set = 23,
Order of Grand Set = 2*Farey(m) - 3 = 43, CPU time in seconds = 0.016`
```

**Program-7.:** This Program in MATHEMATICA is designed to generate the order of the Grand Set of rational numbers, whose numerators and denominators are bounded by Fibonacci(n + 1), using the Euler's Totient Function. The order of this set is 2*Farey(m) − 3, where m = Fibonacci(n + 1). 2*Farey(m) − 3 is the Sequence A176500. The output is shown for n = 5.

```
(*The Grand Set of rational numbers, whose numerators and denominators are bounded by Fibonacci(n + 1)*)
n = 5;
m = Fibonacci[n + 1];
FareyFibonacci = Union[Flatten[Table[a/b, {b, m}, {a, 0, b}]]];
GSet = Complement[FareyFibonacci, {0}];
GrandSet = Union[GSet, 1/GSet];
(*Print[StringForm["n = ``, m = Fibonacci(n + 1) = ``, Farey Fibonacci Set = ``, Order of
Farey Fibonacci Set = ``",
                n, m, FareyFibonacci, Dimensions[FareyFibonacci] ]]*)
(*This line prints the Farey Set with its order*)
Print[StringForm["n = ``, m = Fibonacci(n + 1) = ``, GrandSet = ``, Order of Grand Set =``",
                n, m, GrandSet, Dimensions[GrandSet] ]]
(*This line prints the Grand Set with its oder*)
Print[StringForm["n = ``, m = Fibonacci(n + 1) = ``, Order of Farey Fibonacci Set = ``,
Order of Grand Set = 2*Farey(m) - 3 = ``, CPU time in seconds = ``",
                n, m, Dimensions[FareyFibonacci], Dimensions[GrandSet], TimeUsed[%] ]]
```
$n = 5$, m = Fibonacci(n + 1) = 8, GrandSet =
$\{\frac{1}{8}, \frac{1}{7}, \frac{1}{6}, \frac{1}{5}, \frac{1}{4}, \frac{2}{7}, \frac{1}{3}, \frac{3}{8}, \frac{2}{5}, \frac{3}{7}, \frac{1}{2}, \frac{4}{7}, \frac{3}{5}, \frac{5}{8}, \frac{2}{3}, \frac{5}{7}, \frac{3}{4}, \frac{4}{5}, \frac{5}{6}, \frac{6}{7}, \frac{7}{8}, 1, \frac{8}{7}, \frac{7}{6}, \frac{6}{5},$
$\frac{5}{4}, \frac{4}{3}, \frac{7}{5}, \frac{3}{2}, \frac{8}{5}, \frac{5}{3}, \frac{7}{4}, 2, \frac{7}{3}, \frac{5}{2}, \frac{8}{3}, 3, \frac{7}{2}, 4, 5, 6, 7, 8\}$, Order of Grand Set ={43}

Order of Farey Fibonacci Set = {23}, Order of Grand
  Set = 2*Farey(m) - 3 = {43}, CPU time in seconds = 0.23400000000000043`

**Program-8.:** This Program in MATHEMATICA is designed to generate the Grand Set of rational numbers, whose numerators and denominators are bounded by Fibonacci(n + 1). The order of this set is 2*Farey(m) − 3, where m = Fibonacci(n + 1). 2*Farey(m) − 3 is the Sequence A176500. The output is shown for n = 5.

```
(*The Grand Set of rational numbers in the interval I = [1/n, n], whose numerators and denominators are bounded by Fibonacci(n + 1)*)
n = 5;
m = Fibonacci[n + 1];
FareyFibonacci = Union[Flatten[Table[a/b, {b, m}, {a, 0, b}]]];
FareyInterval = Select[FareyFibonacci, # ≥ 1/n &];
GrandSetInterval = Union[FareyInterval, 1/FareyInterval];
(*Print[StringForm["n = ``, m = Fibonacci(n + 1) = ``, Farey Fibonacci Set = ``, Order of Farey
Set = ``, Farey Fibonacci Set in the Interval I, [1/n, 1] = ``,
                Order of Farey Fibonacci Set in the Interval I, [1/n, 1] = ``",
                n, m, FareyFibonacci, Dimensions[FareyFibonacci], FareyInterval,
Dimensions[FareyInterval] ]]*)
(*This line prints the Farey Fibonacci Set, Farey Fibonacci Set in the Interval with their
orders*)
Print[StringForm["n = ``, m = Fibonacci(n + 1) = ``, Grand Set in the Interval I, [1/n, n] = ``,
Order of Grand Set  in the Interval I, [1/n, n] = ``",
                n, m, GrandSetInterval, Dimensions[GrandSetInterval] ]]
(*This line prints the Grand Set in the Interval with its oder*)
Print[StringForm["n = ``, m = Fibonacci(n + 1) = ``,
                Order of Farey Fibonacci Set in the Interval I, [1/n, 1] = ``, Order of Grand
Set  in the Interval I, [1/n, n] = ``, CPU time in seconds = ``",
                n, m, Dimensions[FareyInterval], Dimensions[GrandSetInterval], TimeUsed[%] ]]
```

n = 5, m = Fibonacci(n + 1) = 8, Grand Set in the Interval

I, [1/n, n] = {$\frac{1}{5}$, $\frac{1}{4}$, $\frac{2}{7}$, $\frac{1}{3}$, $\frac{3}{8}$, $\frac{2}{5}$, $\frac{3}{7}$, $\frac{1}{2}$, $\frac{4}{7}$, $\frac{3}{5}$, $\frac{5}{8}$, $\frac{2}{3}$, $\frac{5}{7}$, $\frac{3}{4}$, $\frac{4}{5}$, $\frac{5}{6}$,

$\frac{6}{7}$, $\frac{7}{8}$, 1, $\frac{8}{7}$, $\frac{7}{6}$, $\frac{6}{5}$, $\frac{5}{4}$, $\frac{4}{3}$, $\frac{7}{5}$, $\frac{3}{2}$, $\frac{8}{5}$, $\frac{5}{3}$, $\frac{7}{4}$, 2, $\frac{7}{3}$, $\frac{5}{2}$, $\frac{8}{3}$, 3, $\frac{7}{2}$, 4, 5},

Order of Farey Fibonacci Set in the Interval I, [1/n, 1] = {19}, Order of Grand

  Set  in the Interval I, [1/n, n] = {37}, CPU time in seconds = 0.5139999999999988`

**Program-9.:** This Program in MATHEMATICA is designed to generate the Grand Set of rational numbers in the interval I = [1/n, n], whose numerators and denominators are both bounded by Fibonacci(n + 1).  The order of this set is the sequence 2*Farey(m; I) – 1, where m = Fibonacci(n + 1).  Farey(Fibonacci(n + 1); I) is the Sequence A176501.  The order of the Grand Set is given by the Sequence A176502.  The output is shown for n = 5.

## Appendix-G

## Integer Sequences

Several integer sequences arise in the present study of the networks formed by equal resistors. Some of these such as the Fibonacci numbers and the Haros-Farey sequence have been well known and extensively studied much before the advent of electrical circuits. The sequence A(n): 1, 2, 4, 9, 22, 53, 131, 337, 869, 2213, … is one of the several sequences occurring in the description of the networks formed by equal resistors.

Here, it is relevant to note the details about *The On-Line Encyclopedia of Integer Sequences* (OEIS), created and maintained by Neil Sloane [4]. The encyclopaedic collection had its genesis in 1965, when Neil J. A. Sloane was looking for the formula for the n-th term of the sequence, 0, 1, 8, 78, 944, 13800, 237432, 4708144, 105822432, 2660215680, 73983185000, 2255828154624, 74841555118992, 2684366717713408, 103512489775594200, 4270718991667353600, 187728592242564421568, 8759085548690928992256, … (now entry A000435 in the OEIS). The last printed edition in 1995 had 5487 sequences, occupying 587 pages. The On-Line Encyclopedia is a ready reference providing comprehensive information and interrelations among the sequences along with bibliography and computer programs for over 176,000 integer sequences. Each sequence has been assigned a unique identity. For instance the sequence A(n) is identified by A048211 and has six additional terms up to n = 22, where as [3] contains the first 16 terms.

As noted earlier, the number of circuit configurations grows much more rapidly than the number of equivalent resistances. The sequence, 1, 2, 4, 10, 24, 66, 180, 522, 1532, 4624, 14136, 43930, 137908, 437502, 1399068, 4507352, 14611576, 47633486, 156047204, 513477502, 1696305728, 5623993944, ..., giving the growth of the number of circuit configurations in the present study, occurs in different contexts such as the number of unlabeled cographs; on n nodes [A000084]. The OEIS enables us to see such connections among unrelated problems. Interestingly enough the OEIS has over a thousand terms of this sequence [A000084]. The constantly updated on-line encyclopedia serves as a treasure house of difficult to compute sequences. In this article we have cited the various sequences by their unique identity occurring in the on-line encyclopedia, and have listed them towards the end of the bibliography.

Appendix-F has nine computer programs using the symbolic package MATHEMATICA, and they just need to be run on a faster computer. Additional terms of the various sequences, obtained can be shared at *The On-Line Encyclopedia of Integer Sequences* [4].

19. **Sequence A005728:** 2, 3, 5, 7, 11, 13, 19, 23, 29, 33, 43, 47, 59, 65, 73, 81, 97, 103, 121, 129, 141, 151, 173, 181, 201, 213, 231, 243, 271, 279, 309, 325, 345, 361, 385, 397, 433, 451, 475, 491, 531, 543, 585, 605, 629, 651, 697, 713, 755, 775, 807, 831, 883, ..., N. J. A. Sloane, **Haros-Farey Sequence**, Sequence **A005728** in N. J. A. Sloane (Editor), *The On-Line Encyclopedia of Integer Sequences* (2008), published electronically at http://www.research.att.com/~njas/sequences/A005728

20. **Sequence A048211:** 1, 2, 4, 9, 22, 53, 131, 337, 869, 2213, 5691, 14517, 37017, 93731, 237465, 601093, 1519815, 3842575, 9720769, 24599577, 62283535, 157807915, ..., Tony Bartoletti (More terms by John W. Layman and Jon E. Schoenfield), **the Number of distinct resistances that can be produced from a circuit of n equal resistors**, Sequence **A048211** in N. J. A. Sloane (Editor), *The On-Line Encyclopedia of Integer Sequences* (2008), published electronically at http://www.research.att.com/~njas/sequences/A048211

21. **Sequence A058351:** 1, 3, 7, 17, 41, 107, 287, 809, 2341, 6965, 21101, 65031, 202939, 640441, 2039509, 6546861, 21158437, 68791923, 224839127, 738316629, 2434622357, 8058616301, ..., N. J. A. Sloane, **Partial Sums of A000084**, Sequence **A058351** in N. J. A. Sloane (Editor), *The On-Line Encyclopedia of Integer Sequences* (2008), published electronically at http://www.research.att.com/~njas/sequences/A058351

22. **Sequence A153588:** 1, 3, 7, 15, 35, 77, 179, 429, 1039, 2525, 6235, 15463, 38513, 96231, 241519, 607339, ..., **Number of resistance values that can be constructed using n 1-ohm resistances by arranging them in an arbitrary series-parallel arrangement**, Sequence **A153588** in N. J. A. Sloane (Editor), *The On-Line Encyclopedia of Integer Sequences* (2008), published electronically at http://www.research.att.com/~njas/sequences/A153588

23. **Sequence A174283:** 1, 2, 4, 9, 23, 57, 151, 409, ..., Sameen Ahmed Khan, **Order of the Set of distinct resistances that can be produced using n equal resistors in, series, parallel and/or bridge configurations**, Sequence **A174283** in N. J. A. Sloane (Editor), *The On-Line Encyclopedia of Integer Sequences* (2008), published electronically at http://www.research.att.com/~njas/sequences/A174283

24. **Sequence A174284:** 1, 3, 7, 15, 35, 79, 193, 489, ..., Sameen Ahmed Khan, **Order of the Set of distinct resistances that can be produced using at most n equal resistors (n or fewer resistors) in series, parallel and/or bridge configurations**, Sequence **A174284** in N. J. A. Sloane (Editor), *The On-Line Encyclopedia of Integer Sequences* (2008), published electronically at http://www.research.att.com/~njas/sequences/A174284

25. **Sequence A174285:** 0, 0, 0, 0, 1, 3, 17, 53, ..., Sameen Ahmed Khan, **Order of the Set of distinct resistances that can be produced using n equal resistors in, series and/or parallel, confined to the five arms (four arms and the diagonal) of a bridge configuration,** Sequence **A174285** in N. J. A. Sloane (Editor), *The On-Line Encyclopedia of Integer Sequences* (2008), published electronically at http://www.research.att.com/~njas/sequences/A174285

26. **Sequence A174286:** 0, 0, 0, 0, 1, 3, 19, 67, ..., Sameen Ahmed Khan, **Order of the Set of distinct resistances that can be produced using at most n equal resistors (n or fewer resistors) in, series and/or parallel, confined to the five arms (four arms and the diagonal) of a bridge configuration,** Sequence **A174286** in N. J. A. Sloane (Editor), *The On-Line Encyclopedia of Integer Sequences* (2008), published electronically at: http://www.research.att.com/~njas/sequences/A174286

27. **Sequence A176497:** 0, 0, 0, 1, 4, 9, 25, 75, 195, 475, 1265, 3135, 7983, 19697, 50003, 126163, 317629, 802945, 2035619, 5158039, 13084381, 33240845, ..., Sameen Ahmed Khan, **Order of the Cross Set which is the subset of the set of distinct resistances that can be produced using n equal resistors in series and/or parallel**, Sequence **A176497** in N. J. A. Sloane (Editor), *The On-Line Encyclopedia of Integer Sequences* (2008), published electronically at: http://www.research.att.com/~njas/sequences/A176497

28. **Sequence A176498:** 0, 0, 0, 0, 0, 0, 0, 0, 1, 6, 9, 24, 58, 124, 312, ..., Sameen Ahmed Khan, **Number of elements less than half in the Cross Set which is the subset of the set of distinct resistances that can be produced using n equal resistors in series and/or parallel**, Sequence **A176498** in N. J. A. Sloane (Editor), *The On-Line Encyclopedia of Integer Sequences* (2008), published electronically at: http://www.research.att.com/~njas/sequences/A176498

29. **Sequence A176499:** 2, 3, 5, 11, 23, 59, 141, 361, 941, 2457, 6331, 16619, 43359, 113159, 296385, 775897, 2030103, 5315385, 13912615, 36421835, 95355147, 249635525, 653525857, 1710966825, 4479358275, ..., Sameen Ahmed Khan, **Haros-Farey Sequence whose argument is the Fibonacci Number; Farey(m) where m = Fibonacci (n + 1)**, Sequence **A176499** in N. J. A. Sloane (Editor), *The On-Line Encyclopedia of Integer Sequences* (2008), published electronically at: http://www.research.att.com/~njas/sequences/A176499

30. **Sequence A176500:** 1, 3, 7, 19, 43, 115, 279, 719, 1879, 4911, 12659, 33235, 86715, 226315, 592767, 1551791, 4060203, 10630767, 27825227, 72843667, 190710291, 499271047, 1307051711, 3421933647, 8958716547, ..., Sameen Ahmed Khan, **2Farey(m) - 3 where m = Fibonacci (n + 1)**, Sequence **A176500** in N. J. A. Sloane (Editor), *The On-Line Encyclopedia of Integer Sequences*

(2008), published electronically at: http://www.research.att.com/~njas/sequences/A176500